# Design of an Instrumentation Unit with Datalogger for Heat Flow Measurement in Solid Metals

Ben Festus[1], Ewetumo T.[2], Adedayo K. D.[2], and Oluyamo S. S.[2]

[1]Physics/Electronics Unit (SLT Department), Federal Polytechnic Ede, Osun State, Nigeria
[2]Physics Department, Federal University of Technology, Akure, Ondo State, Nigeria

**ABSTRACT**

An instrumentation unit that measures heat flow along the test column of solid metal samples is described. The study also describes the design of a datalogger using Arduino Mega Microcontroller. The designed instrumentation unit incorporates current and voltage sensing unit, ten thermocouple sensors and amplifiers, microSD card shield, two LCDs, four microcontrolled switching relays and a 12 V DC water pump motor. The heat flow parameters measured by the instrumentation unit includes temperature gradient, heater current and heater voltage. The designed datalogger logs the measured value of temperature, current and voltage at a time of ten minutes. An area of application for this study is in the development of a device for measurement of thermal conductivity of solid metals. It is an improvement over existing heat flow measurement techniques.

**Keywords:** Datalogger, Microcontroller, Arduino Mega, Arduino 256, Thermocouple, Instrumentation, Heat Flow, Temperature Measurement, Current sensor, Metals

## INTRODUCTION

One of the fundamental techniques for determining heat flow especially in solid metals is by applying a solution of the Fourier law which states that "the heat flux density $\vec{q}$ is proportional to the temperature gradient $T$ in an isotropic body" [1]. The temperature gradient allows heat to flow from a high temperature region to a low temperature region [2] by Fourier law. This statement is expressed mathematically as:

$$\vec{q} = -k\, grad\, T \Rightarrow Q = \frac{d\vec{q}}{dt} = -k.A.\frac{dT}{dx} \qquad (1)$$

Where $Q$ is the amount of heat current, $dx$ is change in length or distance moved, $A$ is unit area, $(dT/dt)$ is temperature gradient, and $k$ is the convective heat transfer coefficient expressed in $(W/m^2 K)$

Generally, energy flow in the form of heat takes place from the hotter, more energetic state, to the colder, less energetic state. This flow of heat has been found not to be just a function of temperature difference alone, but also a function of thermophysical properties, dimensions, time, geometries, and fluids flow [3, 4]. Accurate and precise measurement of the parameters governing heat flow as captured in equation 1 is not only useful but also necessary in order to reduce experimental uncertainties which are very much prevalent in such measurements [5, 6, 7]. This experimental challenge is best addressed using an instrumentation unit with data logging capabilities.

A data logger is an electronic device that automatically records, scans and retrieves measurement data with high speed and greater efficiency over time [8]. Data loggers are small, battery-powered devices that are equipped with a microprocessor, data storage and sensor [9]. A microcontroller-based data logging system requires no additional hardware as data can be collected over specified time intervals and exported to a computer for additional data analysis using appropriate software installed on the computer [10].

In this study, an instrumentation unit with data logging capabilities is designed to measure accurately and record heat flow parameters which are direct consequences of equation 1. The designed instrumentation unit comprises of a microcontroller unit (an Arduino 2560), current sensing unit (ACS712 module), ten type K thermocouples (for temperature measurement), ten thermocouple amplifiers (AD8495 modules), voltage measuring unit, digital displays (two LCDs), four microcontrolled switching relays, and microSD card shield (for recoding measured data). The instrumentation unit designed in this study measures temperature gradient along the test column, current across the heating element and voltage through the heating element. These measured parameters are captured, recorded and stored by the designed microcontroller-based data logging system over a specified time interval. The stored data are then subjected to further analysis in line with the required experimental procedure.

## MATERIALS AND METHODS

Figure 1 shows the basic block diagram adopted in designing the instrumentation unit for this study. The specimen whose heat flow parameters is to be measured is sandwiched between two heater/cooler blocks, such that heat is applied to the specimen from one end (the heater block) and cold water applied from the other end (the cooler block). The fabrication process of the heater/cooler blocks is described in details by [11]. Power measurement is achieved using a designed heat flux measurement schematic in conjunction with a hall-effect linear based current sensing module. Calibrated thermocouple sensors were used to measure the temperature gradient and the resulting voltage reading converted to degree Celsius using a high precision instrumentation amplifier with thermocouple cold junction compensator on an integrated circuit.

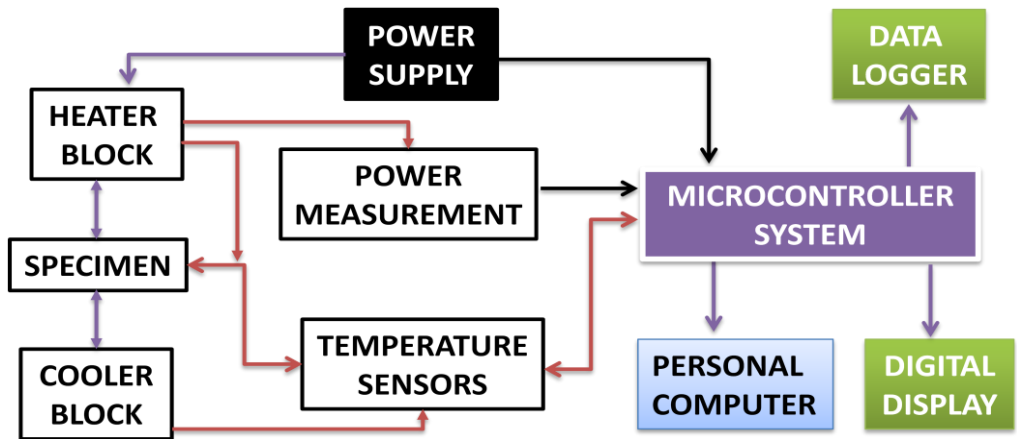

*Figure 1: Basic Block Diagram of the Designed Instrumentation Unit*

Figure 2 shows the experimental setup diagram of the device developed for use with the instrumentation unit. The whole operation of the instrumentation unit is digitally controlled by an Arduino ATmega microcontroller which uses a program written in C++ language to acquire measured data from the different sensors, display the measured data on two liquid crystal digital displays, and subsequently uses the data logger to store them in a microSD card shield over a specified time interval. The heat flow measurement device is designed in accordance with the ASTM E-1225 comparative cut bar test method (TA Instrument, 2010).

Figure 3 shows the power distribution schematic designed using Express PCB software. Voltage was dropped across each section of the components based on their load requirements as shown in the schematic drawn in Figure 3. The 12 V 200 AH Deep Cycle Gel battery used to power the instrumentation unit and the heat flow measurement apparatus was equipped with diode protection to ensure effective and smooth operation of the microcontroller. Relays were also incorporated and digitally controlled by the microcontroller to perform switching application where necessary.

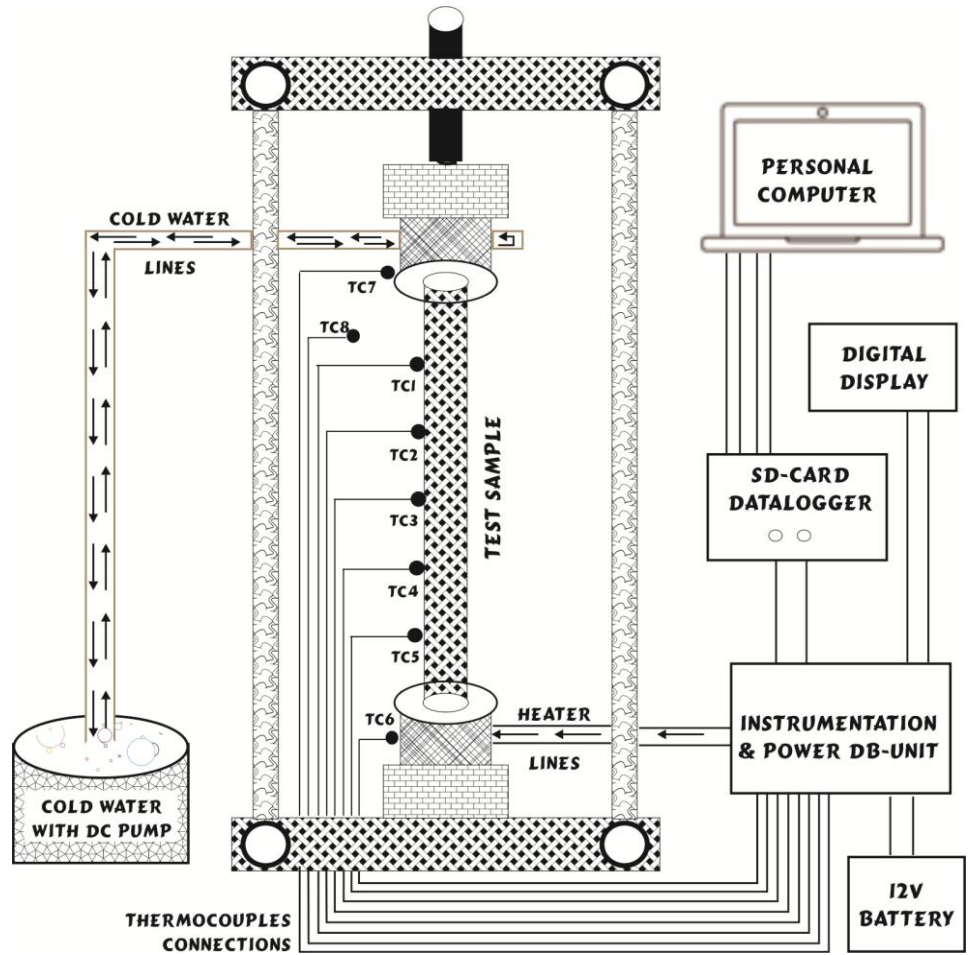

*Figure 2:     Experimental setup of the heat flow measurement device*

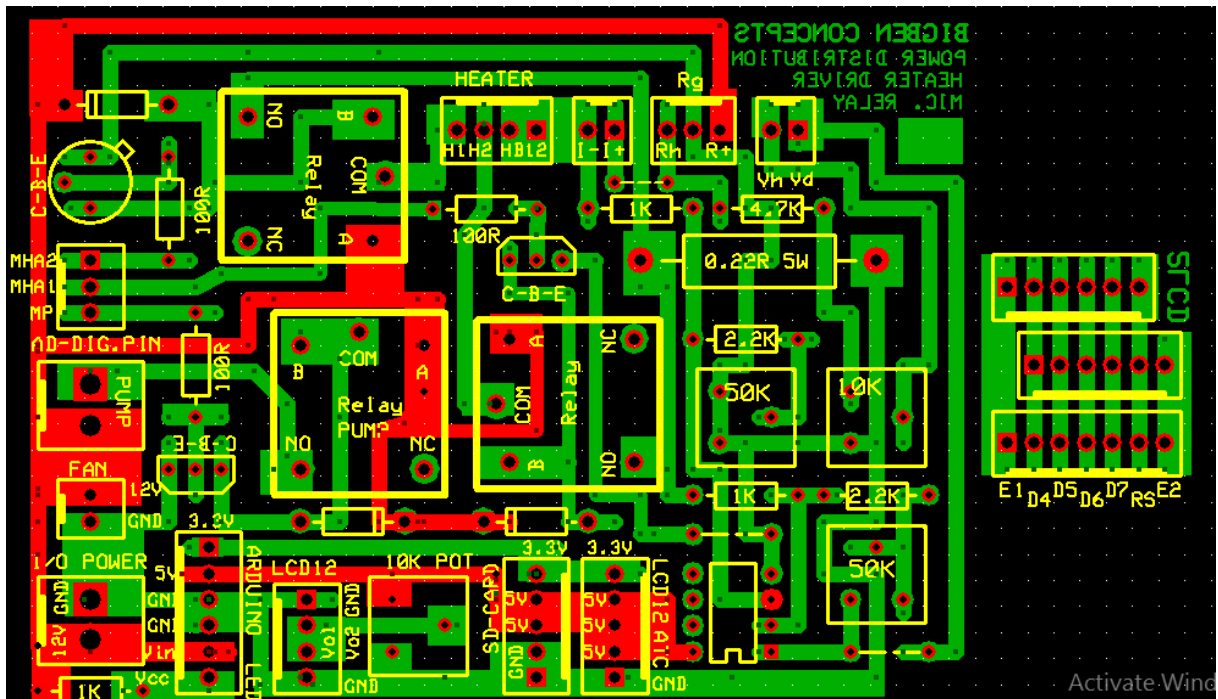

*Figure 3:     PCB designed Microcontroller based Power Distribution schematic*

**Temperature Sensing and Measurement**

The thermo-junction temperature sensing technique was used in this study. Type K thermocouples was used as temperature sensors owing to its ability to offer wide temperature range, low standard error, nearly linear response curve, and its good corrosion resistance [1, 13]. A manually adjustable thermocouple holder is fabricated using wood, plastic, spring, bolt and nuts. This design was necessary to minimize the effect of drift occurring in thermocouple readings and prevent development of high contact resistance (as observed in [7, 14]).

A small hole was drilled into the plastic-wooden holder to hold the thermocouples meant for the test sample in place. When a test sample is fixed in the test column, the thermocouple holder is adjusted until the thermocouples is penned at the appropriate displacement, and after the experiment has been concluded, the heat flux holder is adjusted outwards to allow for the test sample to be removed easily. This design also minimizes the chances of short circuiting the thermocouple wires. The location of temperature sensors in the test column is done symmetrically from top to bottom. In all, ten thermocouples was used in the test column, with two mounted on the cooler block, five mounted along the sample, two mounted at the heater block and one used to monitor the temperature of the chamber. The analogue output of the calibrated thermocouple sensors is connected to the inputs of AD8495 thermocouple amplifier.

Ten AD8495 thermocouple amplifiers were used with the thermocouple sensors to measure the temperature gradient developed during experimental procedures using the schematic shown in Figure 4. Figure 5 shows the schematic of Figure 4 redesigned on a Printed Circuit Board (PCB) using Express PCB software and the electronic components soldered on the PCB as described by the circuit diagram. The temperature sensing unit was calibrated by dissolving some gram of ice cubes in distilled water. The resulting mixture was heated to boiling point with the temperature change and thermocouple potential measured simultaneously by both a liquid-in-glass thermometer (with an uncertainty of $\pm 0.5$ $^o$C) and the AD8495 thermocouple amplifier (with an uncertainty of $\pm 2.0$ $^o$C) for every 5 $^o$C rise in temperature. The absolute deviation was computed and used to appropriately calibrate the AD8495 thermocouple amplifier by modifying the AD8495 conversion equation provided initially by the manufacturer.

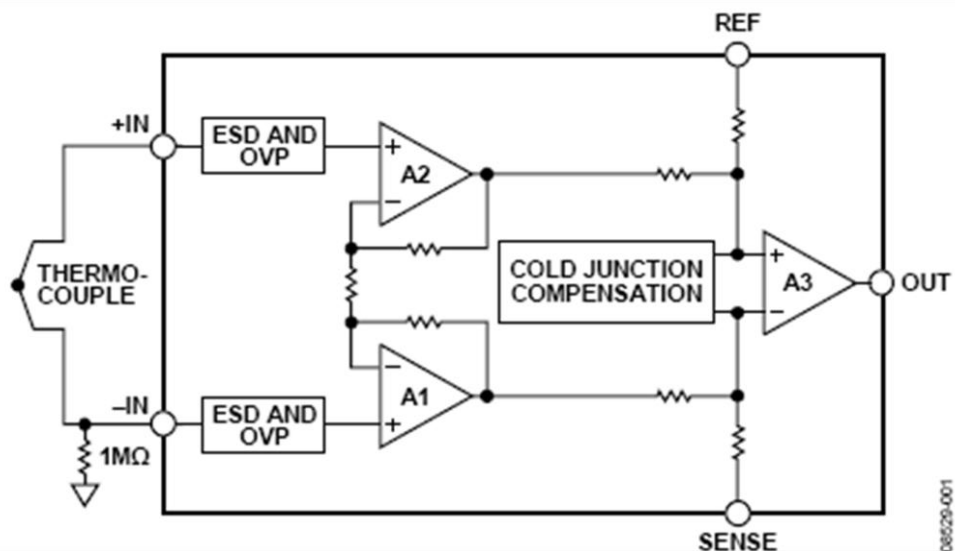

*Figure 4: Heat flux sensor signal conditioner schematic*

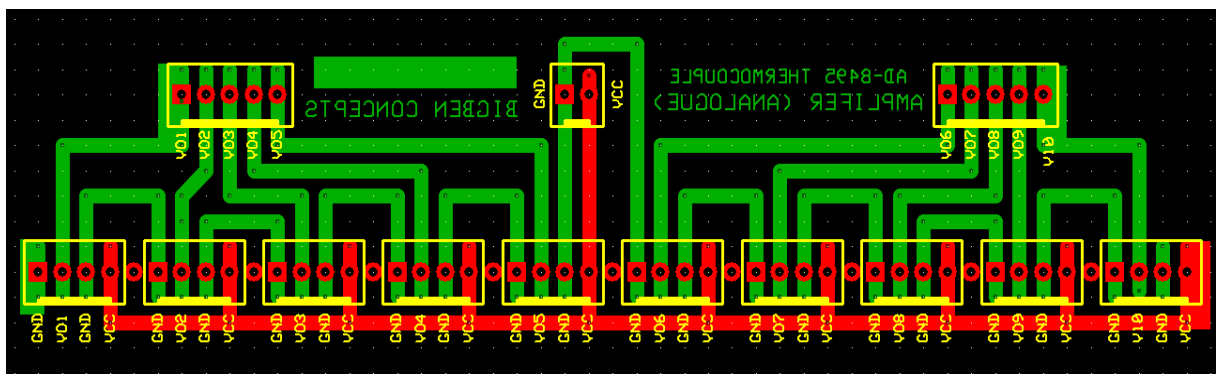

*Figure 5:        PCB designed AD8495 based heat flux measurement circuit*

**Heat Flux Measurement**

To determine the quantity of heat flowing through the solid metal sample, an accurate measurement of the amount of current across the heating element as well as the voltage through the heating element is very essential. Current sensing and measurement was done using ACS712 hall-effect linear current sensor module with a relative uncertainty of $\pm 1.5\%$. The ACS712 was calibrated using a digital multimeter and the result of the calibration used to modify the conversion equation (provided by the manufacturer) embedded inside the microcontroller program.

For voltage sensing and measurement, this was achieved by designing a differential amplifier whose differential voltage gain $A_{Vd}$ is given by equation 2 [15].

$$A_{Vd} = \frac{V_0}{V_2 - V_1} = \frac{R_2}{R_1} \qquad (2)$$

Figure 6 shows the instrumentation circuit designed for measuring the heat flux voltage. By choosing $A_{Vd} = 20$, and setting $R_1 = R_3 = 1K\Omega$ [15], the feedback resistor $R_f$ can be easily determined by applying equation 2 as below:

$$\boldsymbol{R_f = A_{Vd} \times R_1 = 20 \times 1K\Omega = 20K\Omega} \qquad (3)$$

By applying the concept of virtual short, the value of the grounded resistor $R_G$ can be easily determined to yield equation 4:

$$\boldsymbol{R_G = \frac{R_f}{R_1} \times R_3 = \frac{20}{1} \times 1 = 20K\Omega} \qquad (4)$$

With reference to Figure 5, the feedback resistor $R_f = R_{V4} + R_5$ while the grounded resistor $R_G = R_{V3} + R_4$. The differential input resistance $R_i$ is computed to yield equation 5:

$$\boldsymbol{R_i = 2 \times R_1 = 2 \times 1K\Omega = 2K\Omega} \qquad (5)$$

The CMMR(db) is thus computed as given by equation 6

$$\boldsymbol{CMMR(db) = 20 log_{10} \left| \frac{20.0002}{0.0004} \right| = 50 db} \qquad (6)$$

The value obtained for the CMMR from equation 6 shows a high value which is an indication that the differential amplifier designed for measuring the heat flux voltage is a very good design (Donald, 2001). Similarly, since $R_f = R_G = 20K\Omega$, to satisfy design requirements, the proposed value of $R_f$ and $R_G$ was doubled and a variable resistor closest to the new value i.e. $50K\Omega$ was used for $R_{V3}$ and $R_{V4}$. The value of $R_4$ and $R_5$ was randomly selected as $2.2K\Omega$ based on the readily available low value resistors to minimize the effect of adjusting the variable resistor to either its lowest or highest value which most times damages the variable resistor.

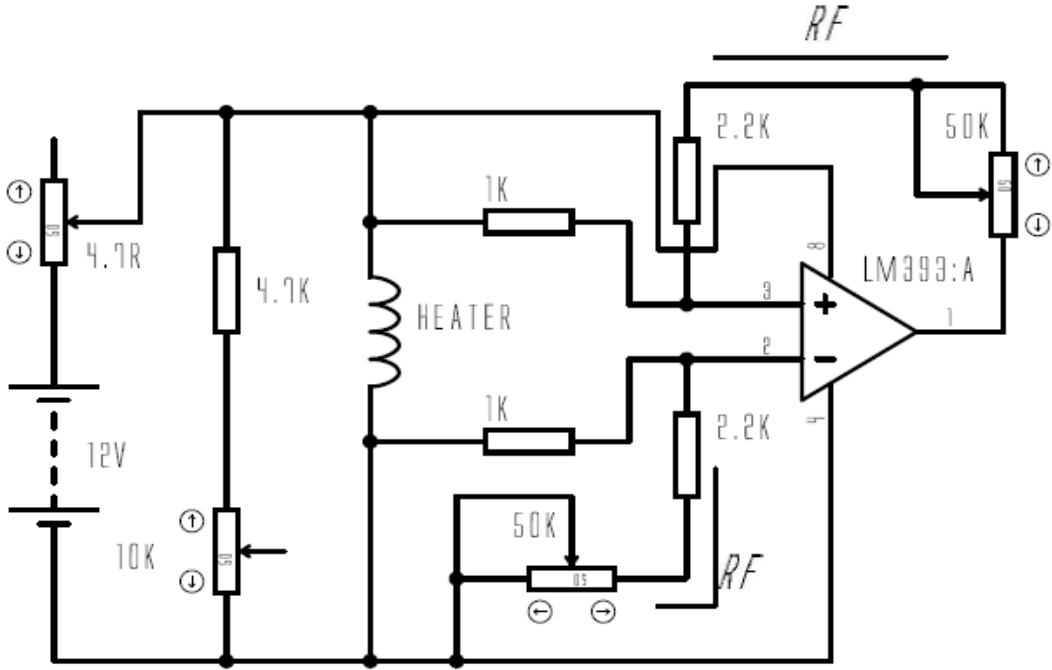

*Figure 6:    Heat flux voltage measurement schematic*

Arduino microcontrollers basically samples between 0 to 5 V as voltage input from external components at a resolution of 1024. To monitor the amount of voltage flowing to the heater using the Ardunio mega 2560 microcontroller, it became necessary to design a small voltage drop circuit that would drop the battery voltage from 12 V to 5 V for use in the microcontroller and calibrate the measurement to obtain a characteristic equation that would be used by the microcontroller. Therefore using the voltage drop equation given, resistors $R_1$ and $R_{V2}$ is carefully selected as a solution to equation 7.

$$V_{IN} = V_{R2} = \frac{R_1}{R_1 + R_{V2}} \cdot V_S \qquad (7)$$

Since the source voltage from the battery is known to be $V_S = 12\,V$, and the expected voltage drop across resistor $R_{V2}$ is also known to be $V_{IN} = V_{R2} = 5\,V$, then, solving for $R_{V2}$ and $R_1$ using equation 7 yields:

$$5R_1 + 5R_{V2} = 12R_1 \qquad (8)$$

Setting $R_1 = 4.7K\Omega$ and using equation 3.12 above, $R_{V2} = 6.58K\Omega$. However to take care of computation errors due to decimal manipulations, a $10K\Omega$ variable resistor was used as $R_{V2}$.

**RESULTS AND DISCUSSION**

The heat flow measurement in the solid metal sample was performed in line with the experimental procedure of the ASTM E-1225 axial heat flow test method. The result of the measurement was acquired by the designed instrumentation unit and logged onto a microSD card by the microcontrolled-based data logger. Figure 7 shows the result of the thermocouple calibration at a 5 $^0$C temperature interval rise. The measured voltage output in mV was converted to $^0$C by the AD8495 thermocouple amplifier. The absolute deviation was determined by subtracting the temperature readings of the liquid-in-glass thermometer from that of the AD8495 output as seen in Table 1. With a percentage deviation of 0.38%, it means that the calibrated thermocouples have a standard measurement error of 0.0038, thus making the sensors highly efficient.

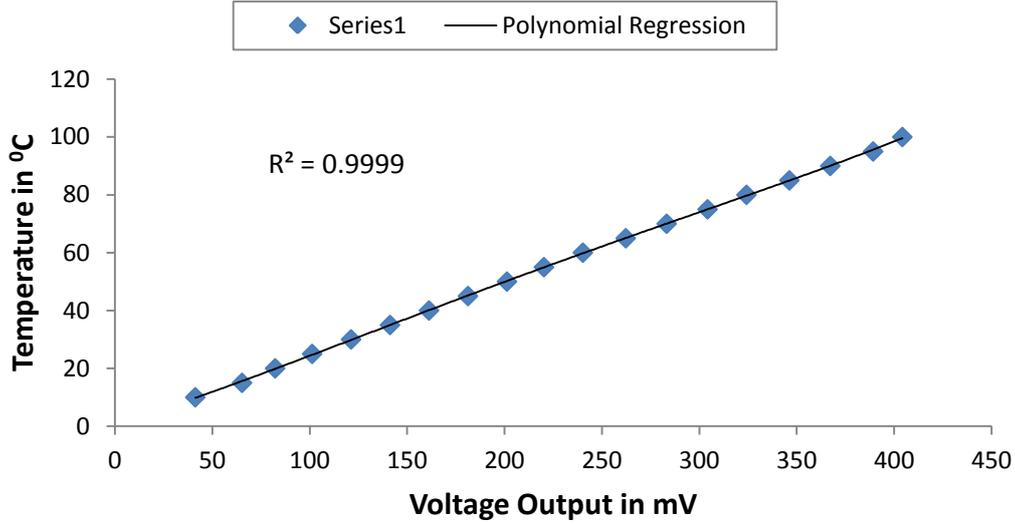

*Figure 7: Calibration Plot for Voltage-to-Temperature Conversion of the Thermocouple*

**Table 1:** Temperature and Voltage Readings of the Thermocouple at an Interval of 5 $^0C$ Rise in Temperature

| Temperature ($^0C$) | Voltage Output (mV)[1] | AD8495 Output ($^0C$) | Absolute deviation ($^0C$) |
|---|---|---|---|
| 10 | 41.203 | 9.8110 | 0.1890 |
| 15 | 65.203 | 15.5971 | 0.5971 |
| 20 | 82.203 | 19.8394 | 0.1606 |
| 25 | 101.203 | 24.6706 | 0.3294 |
| 30 | 121.203 | 29.8117 | 0.1883 |
| 35 | 141.203 | 34.9669 | 0.0331 |
| 40 | 161.203 | 40.1008 | 0.1008 |
| 45 | 181.203 | 45.1867 | 0.1867 |
| 50 | 201.203 | 50.2061 | 0.2061 |
| 55 | 220.203 | 54.9036 | 0.0964 |
| 60 | 240.203 | 59.7719 | 0.2281 |
| 65 | 262.203 | 65.0444 | 0.0444 |
| 70 | 283.203 | 70.0145 | 0.0145 |
| 75 | 304.203 | 74.9518 | 0.0482 |
| 80 | 324.203 | 79.6582 | 0.3418 |
| 85 | 346.203 | 84.8908 | 0.1092 |
| 90 | 367.203 | 89.9998 | 0.0002 |
| 95 | 389.203 | 95.5480 | 0.5480 |
| 100 | 404.203 | 99.4914 | 0.5086 |

The ACS712 current sensor equation embedded in the microcontroller is given by equation 9:

$$Current = \frac{(V_{IOUT} - ACS_{OFFSET})}{Sensitivity} \qquad (9)$$

After calibration of the current sensor, an $ACS_{OFFSET}$ value of $2.47V$ as opposed to $2.5V$ recommended by the manufacturer, the $Sensitivity$ value of $100 \ mV/A$ was used while the $V_{IOUT}$ was obtained using equation 10 based on the microcontroller analog pin readings.

$$clogvolt \ = \ (analogRead(A11) \ / \ 1024.0) \ \times \ 5030 \qquad (10)$$

After all necessary calibrations, Table 2 shows the mean result (over three repeats) of a typical measurement obtained using the designed instrumentation unit for heat flow measurement in a solid metal sample. The unit was switched off when steady state values had been recorded. The whole experiment lasted over three hours and with the aid of the data logger, the temperature, current and voltage parameters were accurately captured, recorded and stored for further analysis. The use of an efficient instrumentation unit like the one designed in this study has greatly helped to eradicate human error which would have occurred if the readings were taking manually over a three hours duration.

**Table 2:** Measured Results for Unknown Solid Sample using Developed Device

| Time (min) | Heater Voltage (V)[1] | Heater Current (A)[1] | Heater Power (W) | T1 ($^0$C)[1] | T2 ($^0$C)[1] | T3 ($^0$C)[1] | T4 ($^0$C)[1] | T5 ($^0$C)[1] | T6 ($^0$C)[1] | T7 ($^0$C)[1] | T8 ($^0$C)[1] |
|---|---|---|---|---|---|---|---|---|---|---|---|
| 10 | 11.90 | 1.08 | 12.85 | 33.47 | 30.54 | 27.61 | 28.59 | 31.52 | 34.45 | 30.54 | 22.72 |
| 20 | 11.90 | 1.08 | 12.85 | 71.60 | 59.87 | 40.32 | 36.41 | 35.43 | 35.43 | 33.47 | 28.59 |
| 30 | 11.90 | 1.08 | 12.85 | 73.55 | 65.73 | 44.23 | 39.34 | 37.38 | 36.41 | 33.47 | 29.56 |
| 40 | 11.90 | 1.08 | 12.85 | 77.46 | 73.55 | 48.14 | 42.27 | 38.36 | 37.38 | 33.47 | 30.54 |
| 50 | 11.90 | 1.08 | 12.85 | 82.35 | 74.53 | 49.11 | 43.25 | 40.32 | 38.36 | 33.47 | 31.52 |
| 60 | 11.90 | 1.08 | 12.85 | 88.26 | 78.44 | 52.05 | 46.18 | 42.27 | 39.34 | 33.47 | 31.52 |
| 70 | 11.90 | 1.08 | 12.85 | 91.15 | 81.37 | 53.02 | 47.16 | 43.25 | 40.32 | 33.47 | 31.52 |
| 80 | 11.90 | 1.08 | 12.85 | 94.08 | 82.35 | 54.00 | 47.16 | 44.23 | 41.29 | 33.47 | 31.52 |
| 90 | 11.90 | 1.08 | 12.85 | 93.10 | 84.30 | 55.96 | 50.09 | 45.20 | 42.27 | 33.47 | 31.52 |
| 100 | 11.90 | 1.08 | 12.85 | 91.15 | 85.28 | 56.94 | 51.07 | 46.18 | 43.25 | 34.45 | 32.50 |
| 110 | 11.90 | 1.08 | 12.85 | 94.08 | 85.28 | 57.91 | 52.05 | 47.16 | 44.23 | 34.45 | 32.50 |
| 120 | 11.88 | 1.08 | 12.83 | 94.08 | 85.28 | 58.89 | 53.02 | 48.14 | 45.20 | 34.45 | 33.47 |
| 130 | 11.88 | 1.08 | 12.83 | 94.08 | 86.26 | 59.87 | 54.00 | 49.11 | 46.18 | 34.45 | 33.47 |
| 140 | 11.88 | 1.08 | 12.83 | 94.08 | 85.28 | 59.87 | 54.98 | 50.09 | 47.16 | 34.45 | 33.47 |
| 150 | 11.88 | 1.08 | 12.83 | 92.12 | 86.26 | 59.87 | 54.98 | 50.09 | 48.14 | 35.43 | 33.47 |
| 160 | 11.88 | 1.08 | 12.83 | 92.12 | 86.26 | 60.84 | 54.98 | 50.09 | 48.14 | 35.43 | 33.47 |
| 170 | 11.88 | 1.08 | 12.83 | 91.15 | 85.28 | 60.84 | 55.96 | 51.07 | 49.11 | 35.43 | 33.47 |
| 180 | 11.88 | 1.08 | 12.83 | 90.17 | 85.28 | 61.82 | 55.96 | 51.07 | 50.09 | 35.43 | 33.47 |
| 190 | 11.88 | 1.08 | 12.83 | 84.30 | 81.37 | 60.84 | 55.96 | 52.05 | 51.07 | 36.41 | 33.47 |
| 200 | 11.88 | 1.08 | 12.83 | 90.17 | 88.21 | 59.87 | 63.78 | 60.84 | 59.87 | 44.23 | 41.29 |
| 210 | 7.12 | 1.80 | 12.85 | 92.12 | 88.21 | 67.69 | 65.73 | 63.45 | 61.82 | 47.16 | 43.90 |
| 220 | 7.12 | 1.80 | 12.85 | 88.87 | 87.88 | 69.31 | 66.06 | 63.78 | 61.82 | 47.16 | 43.57 |
| 230 | 7.12 | 1.80 | 12.85 | 88.22 | 86.91 | 69.31 | 65.73 | 63.78 | 61.82 | 47.16 | 43.57 |
| 240 | 7.07 | 1.80 | 12.76 | 88.86 | 86.26 | 68.66 | 65.73 | 63.78 | 61.82 | 47.16 | 43.25 |
| 250 | 7.07 | 1.80 | 12.76 | 87.24 | 85.93 | 68.66 | 65.73 | 63.78 | 61.82 | 47.16 | 43.25 |
| 260 | 7.07 | 1.80 | 12.76 | 87.24 | 84.95 | 68.66 | 65.73 | 63.78 | 61.82 | 47.16 | 43.25 |
| **270** | **7.07** | **1.80** | **12.76** | **87.24** | **84.30** | **68.66** | **65.73** | **63.78** | **61.82** | **47.16** | **43.25** |
| **280** | **7.07** | **1.80** | **12.76** | **87.24** | **84.30** | **68.66** | **65.73** | **63.78** | **61.82** | **47.16** | **43.25** |
| **290** | **7.07** | **1.80** | **12.76** | **87.24** | **84.30** | **68.66** | **65.73** | **63.78** | **61.82** | **47.16** | **43.25** |
| **300** | **7.07** | **1.80** | **12.76** | **87.24** | **84.30** | **68.66** | **65.73** | **63.78** | **61.82** | **47.16** | **43.25** |

[1] Mean reading taken over 3 repeats
**Note:** *The bold values represent the steady state results*

**Uncertainty Analysis**

This study does not seek to compute the overall uncertainty of the experimental measurement, (this is done in a separate publication), however an attempt is made to estimate the relative uncertainty of each independent measurement and correct likely errors that can affect the measurement precision and accuracy of the instrumentation unit designed in this study. Uncontrolled heat loss has been equally identified as a potential source of measurement error, this was minimized by proper lagging and enclosing the test column in a vacuum tight enclosure using a pyrex glass.

There is every possibility that manual recording of experimental measurements will increase the bias limit which could also affect the precision limit, thereby ultimately introducing high uncertainties in the final measurement. Fixed error was corrected by ensuring proper calibration as discussed earlier. The calibrated values were then used to correct the fixed errors which could have equally affected the overall uncertainty. The relative uncertainty is estimated by [5, 16]:

$$U_R = \pm \left[ \left( \frac{x_1}{R} \frac{\partial R}{\partial x_1} U_1 \right)^2 + \left( \frac{x_2}{R} \frac{\partial R}{\partial x_2} U_2 \right)^2 + \cdots + \left( \frac{x_n}{R} \frac{\partial R}{\partial x_n} U_n \right)^2 \right]^{1/2} \qquad (11)$$

The liquid-in-glass thermometer used for calibrating the thermocouple sensor had a probable measurement error of $\pm 0.5\ ^0C$ while the AD8495 thermocouple amplifier had an initial error of $\pm 1.50\ ^0C$ but which was reduced to $\pm 0.114\ ^0C$ after calibration. Thus the relative uncertainty for the calibrated temperature measurement at $30\ ^0C$ amounts to $\pm 0.0038\ ^0C$. The relative uncertainty of the ACS712 current sensor was also reduced from $\pm 1.5\%$ to $\pm 0.75\%$ after calibration. With increasing heat flux, measurement error decreases appreciably over the specified time interval.

## CONCLUSION

Manually recording experimental measurements especially those of long durations like the one investigated in this study can introduce a lot of random errors which will ultimately affect the overall uncertainty of the test apparatus. To overcome this, this study described the design of an instrumentation unit using well precise measurement modules as well as a microcontroller-based data logger to capture, record and store measurement data over a specified time duration. The data logger ensures that the integrity of the acquired measurement data are preserved over time. All the sensors fitted to the instrumentation unit were properly calibrated to remove fixed errors that could affect the precision and bias limits.

## ACKNOWLEDGEMENT

**The author wish to appreciate TETFUND, and Federal Polytechnic Ede, Osun State Nigeria for sponsoring this research effort.**